\documentclass[preprint,12pt,authoryear]{elsarticle}

\usepackage{amssymb}

\usepackage[OT1]{fontenc}
\usepackage{amsthm,amsmath}
\usepackage[colorlinks,citecolor=blue,urlcolor=blue]{hyperref}
\usepackage{setspace}
\usepackage[authoryear]{natbib}
\usepackage{amssymb}
\usepackage{graphicx}
\usepackage{float}
\usepackage{url} 

\usepackage{color}
\usepackage{multirow}
\usepackage{color}

\newcounter{examplecounter}

\journal{Computational Statistics \& Data Analysis}

\begin{document}

\begin{frontmatter}



\title{Functional archetype and archetypoid analysis \tnoteref{label1}}
\tnotetext[label1]{{The code and data for reproducing the examples  are available at \href{http://www3.uji.es/~epifanio/RESEARCH/faa.rar}{http://www3.uji.es/$\sim$epifanio/RESEARCH/faa.rar}.} }

\author{Irene Epifanio\corref{cor1}}
\address{Dept. Matem\`atiques and Institut de Matem\`atiques i Aplicacions de Castell\'o. Campus del Riu Sec. Universitat Jaume I, 12071 Castell\'o, Spain}
\cortext[cor1]{Tel.: +34 964728390; fax: +34 964728429.}

\ead{epifanio@uji.es}

\begin{abstract}
Archetype and archetypoid analysis can be extended to functional data.
Each function is {approximated by a convex combination} of actual observations (functional archetypoids) or functional archetypes, which are a {convex combination} of observations in the data set. Well-known Canadian temperature data are used to illustrate
the analysis developed.
 Computational methods are proposed for performing these analyses, based on the coefficients of a basis. Unlike a previous attempt to compute functional archetypes, which was only valid for an orthogonal basis, the proposed methodology can be used for any basis. It is computationally less demanding than the simple approach of discretizing the functions. Multivariate functional archetype and archetypoid analysis are also introduced and applied in an interesting problem about the study of human development around the world over the last 50 years.
These tools can contribute to the understanding of a functional data set, as in the {classical} multivariate case.
\end{abstract}


\end{frontmatter}


\section{Introduction}
\label{introduccion}
Archetype analysis (AA) is a statistical technique that seeks to {approximate} data {by} a convex combination of pure or extremal types called archetypes. Archetypes are built as a convex combination of the observations. AA was first introduced by \citet*{Cutler1994}. More recently, archetypoid analysis (ADA) was introduced by \citet*{Vinue15}. Unlike AA, the pure types in ADA are not a mixture {(convex combination)} of observations, but real observations.
It has been shown that human understanding and interpretation of data is made easier when they are represented by their extreme constituents \citep{Davis2010} by the principle of opposites \citep{Thurau12}. In other words, extremes are better than central points for human interpretation.
Their applications have been growing in recent years, especially after the AA algorithm was implemented in the R package {\bf archetypes} (\citet*{Eugster2009}). ADA is available in the R package {\bf Anthropometry} (\citet*{VinueR}). The fields of application include, for instance, market research (\cite{Li2003}, \cite{Porzio2008}, {\cite{Midgley2013}}), biology (\cite{Esposito2012}), genetics (\cite{Morup2013}), sports (\cite{Eugster2012}), industrial engineering (\cite{EpiVinAle,Vinue15}), the evaluation of scientists (\cite{Seiler2013}), astrophysics (\cite{Chan2003}, \cite{Richards2012}), {e-learning (\cite{Theodosiou})}, multi-document summarization (\cite{Canhasi13,Canhasi14}) and different machine learning problems (\cite{Morup2012}, \cite{Stone2002}).

In the seminal paper by \cite{Cutler1994}, one of the illustrative examples worked with functional observations, i.e, data consisting of a set of functions, although they converted them into a matrix by considering a set of values of each curve (after being smoothed) at certain points. Functional data analysis (FDA) {comprises} statistical procedures for functional
observations (a whole function is a datum).
The  objectives of FDA are essentially the same as those of any
other branch of statistics. Although FDA is relatively new field, some classic references in this field include: \citet{Ramsay05}, who provide an excellent overview, \citet{Ferraty06}, with new methodologies for studying functional data nonparametrically, and \citet{Ramsay02}, who offer interesting applications in different fields, and \cite{Ramsay09} regarding software in this field. More recent advances and interesting applications of FDA comprise a variety of fields such as aviation safety  \citep{Gregorutti201515}, chromatography \citep{Raket2014227},  the quality of cookies and the relationship with the flour kneading process \citep{Jacques201492}, the relationship between the geometry of the internal carotid artery and the presence or absence of an aneurysm \citep{Usset2015,Sangalli09} and
the analysis of hippocampal differences in
Alzheimer's disease \citep{Epifanio11,Epifanio14}.

A first attempt to extend AA to functional data was made by \citet{Constatini2012}. Functions were expressed in a functional basis, and {standard} multivariate AA was applied to the coefficients in this basis. {This method is only valid when the basis is orthonormal. The same thing is true when computing standard principal component analysis (PCA) of the basis coefficients in order to carry out functional principal component analysis (FPCA).} In this paper, {a  methodology is developed} for obtaining functional archetypes and archetypoids, whatever the basis used for approximating the functions.

Interest in describing and displaying the important features of a set of curves is not recent. \citet{JonesRice}  considered curves with extreme scores of principal components. This could be viewed as searching the archetypoid functions. However, unlike PCA, the goal of AA is to obtain extreme individuals, and curves with extreme PCA scores do not necessarily return {archetypal} observations. This is explained in \cite{Cutler1994} and shown in \cite{EpiVinAle} through a problem where  archetypes could not be recovered with PCA even if all the components
had been considered.

In this paper, AA and ADA are extended to univariate and multivariate functional data (more than one function is available per individual). Section \ref{methodology} presents a review of archetype and archetypoid analysis in the {classical} multivariate case, and their respective extensions to FDA are introduced and illustrated through a well-studied data set in the field of FDA. The location of functional archetypes and archetypoids is also analyzed. Human development is a topic of considerable political and public interest, suffice it to say that all United Nations member states committed to help achieve the Millennium Development Goals established in 2000, by 2015 \citep{onu}. In Section \ref{aplicaciones}, {the} proposal is applied to understanding statistics about human development for all countries, in order to obtain the big picture of global development. This can make it easier to  interpret the large amount of data about sustainable development even
 for non-experts. The code in R (\cite{R}) and data for reproducing the results are available at \href{http://www3.uji.es/~epifanio/RESEARCH/faa.rar}{http://www3.uji.es/$\sim$epifanio/RESEARCH/faa.rar}. Conclusions and future work are discussed in Section \ref{conclusiones}.

\section{Definition of functional AA and ADA}
\label{methodology}
\subsection{AA and ADA for {(standard)} multivariate data}\label{AAmult}
Let $\bold{X}$ be an $n \times m$ matrix that contains a {usual} multivariate dataset with $n$ observations and $m$ variables. The objective of AA is to find a $k \times m$ matrix $\bold{Z}$, {whose rows} are the $k$ archetypes in those data, in such a way that data can be {approximated by} mixtures of {the} archetypes. To obtain the archetypes, AA computes two  matrices $\bold{\alpha}$ and $\bold{\beta}$ which minimize the residual sum of squares (RSS) {that arises from combining the equation where $\bold{x}_i$ is approximated by a mixture of $\bold{z}_j$'s (archetypes) {($\sum_{i=1}^n \| \bold{x}_i - \sum_{j=1}^k \alpha_{ij} \bold{z}_j\|^2$)}  and the equation where $\bold{z}_j$'s is expressed as a mixture of the data {($\bold{z}_j = \sum_{l=1}^n \beta_{jl} \bold{x}_l$)}:}

{
\begin{equation} \label{RSSar}
RSS = \displaystyle \sum_{i=1}^n \| \bold{x}_i - \sum_{j=1}^k \alpha_{ij} \bold{z}_j\|^2 = \sum_{i=1}^n \| \bold{x}_i - \sum_{j=1}^k \alpha_{ij} \sum_{l=1}^n \beta_{jl} \bold{x}_l\|^2{,}
\end{equation}
}

under the constraints

\begin{enumerate}

\item[1)] $\displaystyle \sum_{j=1}^k \alpha_{ij} = 1$ with $\alpha_{ij} \geq 0$ {for} $i=1,\ldots,n$ {and}

\item[2)] $\displaystyle \sum_{l=1}^n \beta_{jl} = 1$ with $\beta_{jl} \geq 0$ {for} $j=1,\ldots,k${.}

\end{enumerate}

Constraint 1) means that the {approximations} of $\bold{x}_i$ are a convex combination of archetypes, $\bold{\hat{x}}_i = \displaystyle \sum_{j=1}^k \alpha_{ij} \bold{z}_j$. Each $\alpha_{ij}$ is the weight of the archetype $j$ for the observation $i$; that is to say, the $\alpha$ coefficients indicate how much each archetype contributes to the approximation of each observation. Constraint 2) means that archetypes $\bold{z}_j$ are a mixture of the observations, $\bold{z}_j = \displaystyle \sum_{l=1}^n \beta_{jl} \bold{x}_l$.

Note that archetypes are not necessarily {actual observations}. This would happen if only one $\beta_{jl}$ is equal to $1$ in constraint 2) for each $j$. This implies that $\beta_{jl}$ can only take on the values 0 or 1, since $\beta_{jl} \geq 0$ and the sum of constraint 2) is 1. In ADA, the continuous optimization problem of AA transforms into the following mixed-integer optimization problem:

\begin{equation}\label{arche_adapt}
RSS = \displaystyle \sum_{i=1}^n \| \bold{x}_i - \sum_{j=1}^k \alpha_{ij} \bold{z}_j\|^2 = \sum_{i=1}^n \| \bold{x}_i - \sum_{j=1}^k \alpha_{ij} \sum_{l=1}^n \beta_{jl} \bold{x}_l\|^2 {,}
\end{equation}

under the constraints

\begin{enumerate}

\item[1)] $\displaystyle \sum_{j=1}^k \alpha_{ij} = 1$ with $\alpha_{ij} \geq 0$ and $i=1,\ldots,n$ {and}

{\item[2)] $\displaystyle \sum_{l=1}^n \beta_{jl} = 1$ with $\beta_{jl} \in \{0,1\}$ and $j=1,\ldots,k$.}

\end{enumerate}

{Note that 2) implies that $\beta_{jl}=1$ \mbox{for one and only one $l$} and $\beta_{jl}=0$ otherwise.}

Archetypes and archetypoids are extremal (pure types) representatives of the data. Archetypes belong to the boundary of the convex hull of data if $k$ $>$ 1 (see \cite{Cutler1994}), whereas archetypoids do not necessarily (see \cite{Vinue15}). For $k$ = 1, the archetype is the mean, and the archetypoid is the medoid (with one cluster) (\citet*{Kaufman90}).

In order to solve AA, \citet{Cutler1994} proposed an alternating minimizing algorithm, which was implemented in R by \cite{Eugster2009}. To solve  the convex least squares problems, they used  a penalized version of
the non-negative least squares algorithm by \citet*{Lawson74}. As explained by \cite{Eugster2009}, the problems to solve are of the  $\| \bf{u} - \bf{Tw} \|$ kind, where $\bf{u}$ and $\bf{w}$ are
vectors and $\bf{T}$ is a matrix, all of appropriate dimensions, and with the non-negativity and equality restrictions. In the penalized version an extra element $H$ {(for ``huge'')} is added to $\bf{u}$ and to each observation of $\bf{T}$, minimizing under non-negativity constraints $\| \bf{u} - \bf{Tw} \|$ + $H \| 1 - \bf{w} \|$. The larger $H$ is, the more dominant the second term is, which means that the equality restrictions are fulfilled.
The penalized non-negative least squares method is quite slow, but it is attractive since it can be used if the number of variables
is larger than the number of cases (according to \cite{Cutler1994}).
Note that in the R library {\bf archetypes}, data are standardized by default. 

 To solve ADA, \cite{Vinue15} proposed an algorithm, which was implemented in R by \cite{VinueR}, based on the Partitioning Around Medoids (PAM) clustering algorithm (\cite{Kaufman90}).
That algorithm consists of two phases{, a BUILD step and a SWAP step. An initial set of archetypoids is computed in the BUILD phase}. The SWAP step seeks to improve the set of archetypoids by exchanging chosen observations for unselected cases and by checking if these replacements reduce the RSS.
In the R implementation, the initial candidates are determined in three different ways. The first are the nearest observations in Euclidean distance to the $k$ archetypes, the so-called $cand_{ns}$ set. The second candidates, referred to as the  {$cand_{\alpha}$} set, are the observations with the maximum $\alpha$ value for each archetype, i.e., the observations with the largest relative share for the respective archetype. The third initial candidates, the $cand_{\beta}$ set, consist of the cases with the maximum $\beta$ value for each archetype, i.e., those whose contribution to the generation of archetypes is largest.

Neither archetypes nor archetypoids are necessarily nested. In order to select the value $k$ the user can decide how many representatives are to be considered or the elbow criterion can be used as in \cite{Cutler1994,Eugster2009,Vinue15} (the value $k$ is chosen as the point where the elbow on the RSS representation for a series of different $k$ values is found).

\subsection{AA and ADA for functional data}\label{FAA}
The first consideration is that in the functional context, the values of the $m$ variables in the {standard} multivariate context are replaced by function values with a continuous index $t$. Similarly, summations are replaced by integration to define the inner product.
Let $\{x_1(t),$
$\dots,$ $x_n(t)\}$ be a set of univariate observed functions with argument $t$ defined in the interval $[a,b]$. It is always assumed that these functions satisfy reasonable
smoothness conditions and are square-integrable functions on
that interval, a Hilbert space.
The objective of functional archetype analysis (FAA) is to find $k$ archetype functions, in such a way that our  functional data sample can be {approximated by} mixtures of those archetypes. Analogously, FAA computes two matrices $\bold{\alpha}$ and $\bold{\beta}$ which minimize the residual sum of squares (RSS) as in equation \ref{RSSar}, taking into account that now $\| . \|$ stands for a functional norm instead of a vector norm and that the vectors $\bold{x}_i$ and $\bold{z}_j$ correspond to the functions ${x_i}$ and ${z_j}$. The meaning of $\bold{\alpha}$ and $\bold{\beta}$ in the functional case is identical to the {standard} multivariate case.

In the same manner, functional archetypoid analysis (FADA) seeks $k$ functions of the sample (archetypoids), in such a way that our  functional data sample can be {approximated by} mixtures of those archetypoids. As before, the vector norms are replaced by functional norms in equation \ref{arche_adapt}, and the interpretation is identical, changing only vectors for functions.

The same proofs developed by \cite{Cutler1994} can be used to demonstrate that the functional archetype for $k$ = 1 is the functional mean, whereas for $k$ $>$ 1, functional archetypes belong to the boundary of the convex hull generated by the set $S$ = $\{x_1(t),$
$\dots,$ $x_n(t)\}$, $conv(S)$. {The medoid
is the object in the cluster for which the average dissimilarity to all the objects in the cluster is minimal \citep{Kaufman90}.}
In the archetypoid case, for $k$ = 1, as in the {classical} multivariate case, {the} archetypoid would be the functional medoid (with one cluster) of $\{x_1(t),$
$\dots,$ $x_n(t)\}${. Note that }
 the minimization of RSS coincides with the definition of functional medoid, if the vector norms are replaced by functional norms in its definition. {A vertex of the convex hull of $S$ is a case $x_i$ of $S$ for which $x_i$ does not belong to $conv(S \backslash \{x_i\})$. } For $k$ $>$ 1, archetypoids are not necessarily vertices in the {classical} multivariate case or in the functional case. For example, let us consider the following functions defined in $[0, 1]$, $x_1(t)$ = 0, $x_2(t)$ = 1 if $t$ in $[0, 0.5]$ and 0.8 in $(0.5, 1]$, $x_3(t)$ = 0.8 if $t$ in $[0, 0.5]$ and 1 in $(0.5, 1]$ and $x_4(t)$ = 0.9. For $k$ = 2, the functional archetypoids are $x_1$ and $x_4$, and $x_4$ is not a vertex, since $x_4$ = 0.5$x_2$ + 0.5$x_3$. {The same examples used by \cite{Vinue15} in the standard multivariate case can also be used to show that functional archetypoids are not necessarily on the boundary of the convex hull. 
 The example functions could be a linear combination of (orthonormal) Fourier basis functions, whose coefficients would be the values given in the examples in \cite{Vinue15}}.


\subsubsection{Computational details}
After defining the problem, computational methods for FAA and FADA have to be determined. The $L^2$-norm ($\|f\|^2= <f,f> = \int_a^b f(t)^2 dt$) is considered for RSS computation. Note that, in practice, the functions are not observed continuously,
but rather in a finite set of points. Suppose that preliminary steps such as smoothing have been carried out. A naive approach is to discretize the observed functions to a fine grid of $m$ equally spaced values from $a$ to $b$. This gives an $n \times m$ $\bold{X}$ matrix that can be used with standard multivariate AA and ADA. This was the strategy followed by \cite{Cutler1994} in their seminal paper for working with functional observations. Depending on the features of the functions, the number $m$ may have to be large to capture them, which increases the computational time (in the computational complexity study by \cite{Eugster2009}, increasing the number of variables implied a polynomial increase of the
computation time per iteration in AA). It should also be noted that this approach uses a fairly crude approximation of the integral $\int_a^b f(t)^2 dt$ as the simple sum of discrete values. It would be similar to the trapezoidal rule; in fact, both methods are the same if periodic boundary conditions are considered. Other{,} more sophisticated numerical integration techniques could be used to obtain a higher accuracy for approximating the integrals, at the expense of increasing the computational complexity. {Instead a} basis function expansion of the functions will be used. In this way, a high accuracy will be kept without increasing the computational cost. In fact, as the number of basis functions used is usually smaller than the number of sampling points, the computational time will be decreased with respect to the strategy of discretizing the functions. {This is really important if we work with large scale data with many observations and variables. Note that many relevant problems, where AA has been applied, involve functions with many sampling points, for example, in neurology (\cite{Morup2012}, \cite{Tsanousa20158454}), chemistry \citep{Morup2012}, remote sensing \citep{Zhao}, navigation scenarios \citep{Feld15}, meteorology \citep{climate}, sustainability \citep{Thurau12}, image analysis \citep{Thurau09}, etc.}

 {Representing functions by basis functions has the advantage that it can be applied to data where the
functions have not all been measured at the same time points, data observations do
not have to be equally spaced and the number of sampling points can vary
across cases. The critical point is to select an appropriate basis and the number of basis functions (less computation is required for smaller number of basis functions). This is a common question in all FDA problems. Ideal basis functions should have features that match those known to belong to the functions being approximated (see \cite{Ramsay05} for a good explanation about smoothing functional data). In the classic FDA situation considered here, functions are densely observed and the basis coefficients are estimated separately to each function. With sparsely observed functions, the
 information from all functions should be utilized to estimate the coefficients for each function \citep{James}.}

 {In the basis approach}, each function $x_i$ is expressed as a linear combination of known basis functions $B_h$ with $h$ = 1, ..., $m$: $x_i(t)$ = $\sum_{h=1}^m b_i^h B_h(t)$ = ${\bold{b_i'B}}$, where $'$ stands for transpose and $\bold{b_i}$  indicates the vector of length $m$ of the coefficients and $\bold{B}$ the functional vector whose elements are the basis functions. Along with this, RSS can be expressed (with the corresponding constraints for FAA and FADA) as:

\begin{equation}\label{RSSfar}
\begin{split}
 RSS = \displaystyle \sum_{i=1}^n \| {x}_i - \sum_{j=1}^k \alpha_{ij} {z}_j\|^2 = \sum_{i=1}^n \| {x}_i - \sum_{j=1}^k \alpha_{ij} \sum_{l=1}^n \beta_{jl} {x}_l\|^2 = \\
\sum_{i=1}^n \| \bold{b'}_i\bold{B} - \sum_{j=1}^k \alpha_{ij} \sum_{l=1}^n \beta_{jl} \bold{b'}_l \bold{B}\|^2 =
 \sum_{i=1}^n \| (\bold{b'}_i - \sum_{j=1}^k \alpha_{ij} \sum_{l=1}^n \beta_{jl} \bold{b'}_l) \bold{B}\|^2 =\\  \sum_{i=1}^n \| \bold{a'}_i \bold{B}\|^2 = \sum_{i=1}^n <\bold{a'}_i \bold{B}, \bold{a'}_i \bold{B}> = 
 \sum_{i=1}^n \bold{a'}_i \bold{W} \bold{a}_i
 {,}
\end{split}
\end{equation}
where $\bold{a'}_i$ = $\bold{b'}_i - \sum_{j=1}^k \alpha_{ij} \sum_{l=1}^n \beta_{jl} \bold{b'}_l$ and $\bold{W}$ is the order $m$ symmetric matrix with elements $w_{m_1,m_2}$ = $\int B_{m_1}B_{m_2}$, i.e.  the matrix containing the inner products of the pairs of basis functions. In the case of an orthonormal basis such as Fourier, $\bold{W}$ is the order $m$ identity matrix, and
FAA (FADA, respectively) is reduced to AA (ADA, respectively) of the basis coefficients. But, in other cases, we may have to resort to numerical integration to evaluate $\bold{W}$, but once $\bold{W}$ is computed, no more numerical integrations are necessary.

As we can see, the first attempt to define FAA made by \cite{Constatini2012} is only valid if the basis is orthonormal, but not otherwise.

\begin{figure}[H]
\centering
\begin{tabular}{c}
\includegraphics[width=.8\textwidth]{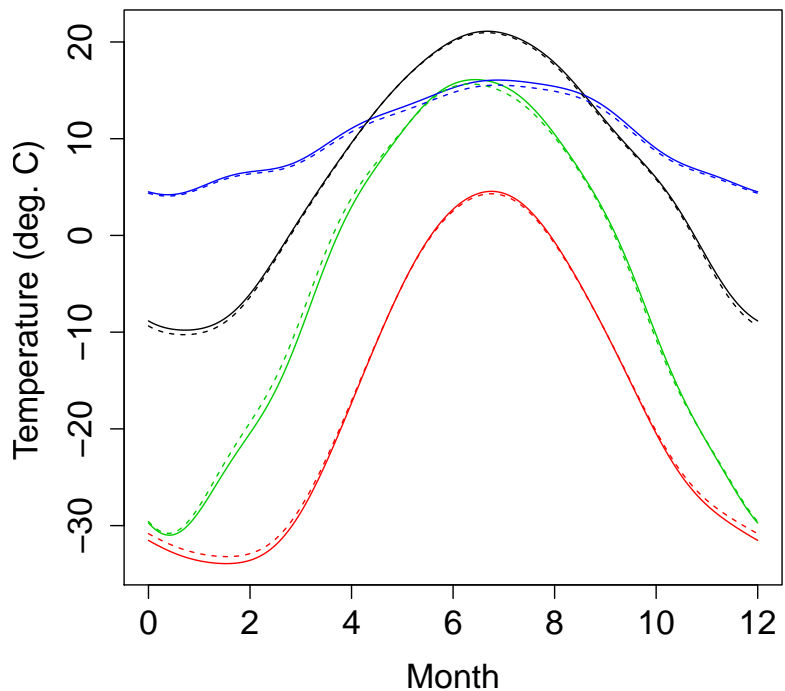}
\end{tabular}
\caption{Four functional archetypes (solid lines) and archetypoids (dashed lines) for Canadian weather stations.\label{canada}}
\end{figure}

To illustrate the concepts, a database heavily analyzed in FDA that appeared in \cite{Ramsay05} to illustrate FPCA is used, so that readers who are interested can appreciate the differences in the analysis. The  mean monthly temperatures for 35 Canadian weather stations averaged over 1960 to 1994 are considered. As in \cite{Ramsay05}, the data are approximated with a 12-term Fourier series. 
As there are four climate zones, four archetypes and archetypoids are chosen. Fig. \ref{canada} shows the solution of FAA and FADA (the same archetypoids are obtained from the three initial candidates). The FADA solution corresponds to the following weather station (each of them belongs to a different climate zone): Montreal (Atlantic), Resolute (Arctic), Dawson (Continental) and Victoria (Pacific), in black, red, green and blue, respectively. If we bring to mind a map of Canada, these stations are located near the borders (extreme zones) of the country.

\subsection{Multivariate FAA and FADA}
We often wish to study more than one function at the same time. For the shake of clarity, the extension of FAA and FADA to deal
with bivariate functional data is discussed. The key is to define an inner product between bivariate
functions, which is computed simply as the sum of the inner products of the two components.
Therefore, the squared norm of a bivariate function is simply
the sum of the squared norms of the two component functions.  What all this amounts to is that FAA or FADA for $M$ multivariate functions is equivalent to $M$ independent FAA or FADA, respectively, with shared parameters $\bold{\alpha}$ and $\bold{\beta}$. In practice, a composite function is formed by
 stringing the $M$ functions
together.

Let $f_i(t)$ = $(x_i(t), y_i(t))$ be a bivariate function. Its squared  norm  is $\|f_i\|^2= \int_a^b x_i(t)^2 dt + \int_a^b y_i(t)^2 dt$. To compute FAA and FADA, let us consider $\bold{b^x_i}$ and  $\bold{b^y_i}$, which are the vectors of length $m$ of the coefficients for $x_i$ and $y_i$ respectively for the basis functions $B_h$. Therefore,

\begin{equation}\label{RSSfarb}
\begin{split}
 RSS = \displaystyle \sum_{i=1}^n \| {f}_i - \sum_{j=1}^k \alpha_{ij} {z}_j\|^2 = \sum_{i=1}^n \| {f}_i - \sum_{j=1}^k \alpha_{ij} \sum_{l=1}^n \beta_{jl} {f}_l\|^2 = \\
 \sum_{i=1}^n \| {x}_i - \sum_{j=1}^k \alpha_{ij} \sum_{l=1}^n \beta_{jl} {x}_l\|^2 + \sum_{i=1}^n \| {y}_i - \sum_{j=1}^k \alpha_{ij} \sum_{l=1}^n \beta_{jl} {y}_l\|^2 = \\ \sum_{i=1}^n \bold{a^x}_i' \bold{W} \bold{a^x}_i + \sum_{i=1}^n \bold{a^y}_i' \bold{W} \bold{a^y}_i
 {,}
\end{split}
\end{equation}
where $\bold{a^x}_i'$ = $\bold{b^x}_i' - \sum_{j=1}^k \alpha_{ij} \sum_{l=1}^n \beta_{jl} \bold{b^x}_l'$ and $\bold{a^y}_i'$ = $\bold{b^y}_i' - \sum_{j=1}^k \alpha_{ij} \sum_{l=1}^n \beta_{jl} \bold{b^y}_l'$, with the corresponding constraints for $\bold{\alpha}$ and $\bold{\beta}$. As before, a penalized version of
the non-negative least squares algorithm is used to solve the minimization, but note that  the observations are now formed by joining
$\bold{b^x_i}$ and  $\bold{b^y_i}$. If the basis is orthonormal, FAA and FADA can be computed by applying the {classical} multivariate version to the $n \times 2m$ coefficient matrix composed by joining the coefficient matrix for $x$ and $y$ components.


\section{Applications and results}
\label{aplicaciones} World Bank Open Data is a free and open access database about development in countries around the globe. Two indicators from the database World Development Indicators \citep{wbdata} are considered here. These indicators are the same as those selected by Rosling in one of his legendary presentations: ``The best stats you've ever seen'' in TED Talks \citep{RoslingTED}, which dispelled myths about the world.
One of the  indicators considered is  total fertility rate (births per woman). Total fertility rate (TFR) represents the number of children that would be born to a woman if she were to live to the end of her childbearing years and bear children in accordance with current age-specific fertility rates.
The other is life expectancy at birth (LEB),  i.e.,  the number of years a newborn infant would live if prevailing patterns of mortality at the time of its birth were to stay the same throughout its life.
The series of each country goes from 1960 to 2013.

Instead of using the motion charts developed by Rosling, I consider the use of functional archetypoids for representing the big picture of global development. As data are {approximated by} a convex combination of the archetypoids, this encourages interpretation in contrast with techniques that express data as a linear combination (not restricted to be between 0 and 1) of certain important latent components \citep{Thurau12}. Archetypoids and archetypes accommodates human cognition, by focusing
on extreme opposites \citep{Thurau12}.
Using functional archetypoids instead of functional archetypes facilitates an intuitive understanding of the results even for non-experts \citep{Vinue15,Thurau12}, as FADA {approximates} the data {by} mixtures of extreme
countries, and not as mixtures of mixtures, as FAA does.

 For some small countries (many of them small islands) data are missing for the majority of years. Only information for the last few years is available. These countries have not been considered. Three countries with missing values for only some of the years have not been erased from the database. In total 190 countries have been considered. All functions (those with or without missing years) are expressed with 32 B-spline basis functions of order 4 (cubic splines) from 1960 to 2013, with equally spaced knots. Having a different number of argument values is not a problem with FDA.

Note that TFR and LEB  
 are measured
in non-compatible units, {so} each functional variable should be standardized. Functional means and variances are defined point wisely across replications \citep{Ramsay05}.  The averages are subtracted from the respective functions, then functions are divided by the respective standard deviation functions. Computationally, this can be done by standardizing the coefficients in the basis. Bivariate FADA is computed after standardization.

\begin{figure}[h]
\centering
\begin{tabular}{cc}
\includegraphics[width=0.5\textwidth]{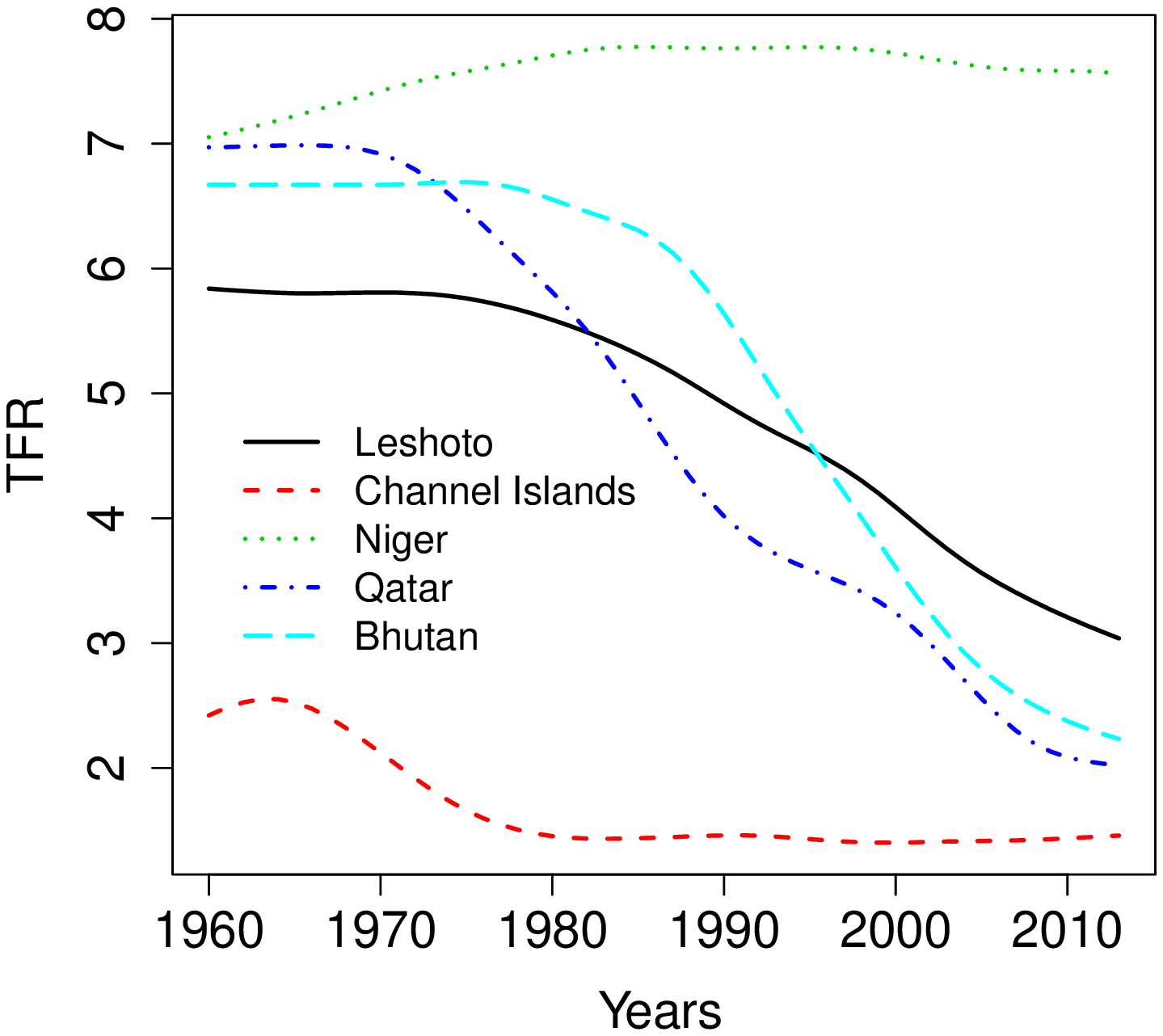} & \includegraphics[width=0.5\textwidth]{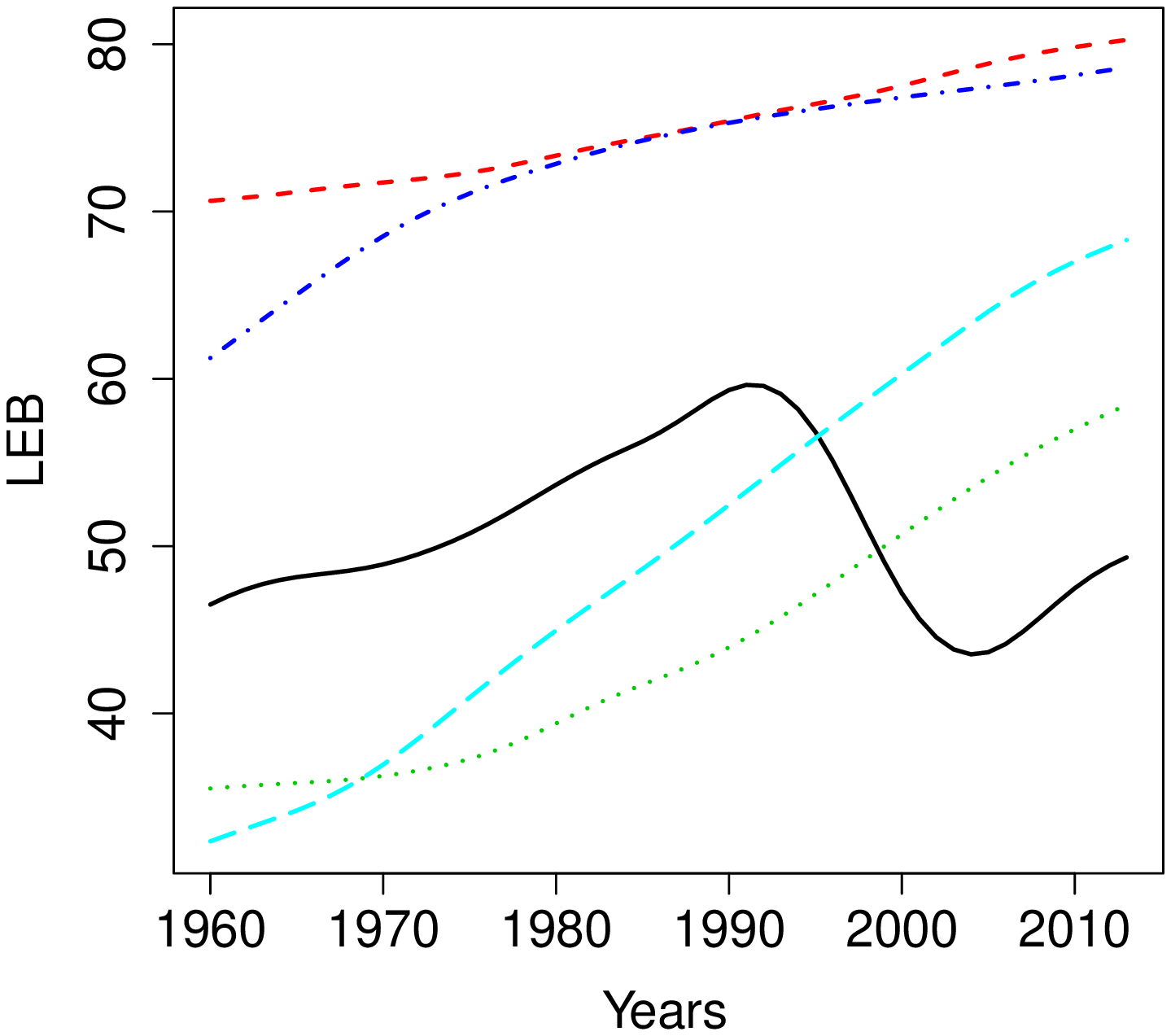}
\end{tabular}
\caption{{Total fertility rate (TFR, left)  and life expectancy at birth (LEB, right)} archetypoid functions for $k$ = 5.\label{arworld}}
\end{figure}

In the interests of brevity
and as an illustrative example we examine the results of 5 functional archetypoids{.} 
Functional archetypoids (the same solution is obtained from the three initial candidates {described in Section \ref{AAmult}}) are (the continent to which each country belongs appears between parentheses): 1) Lesotho (Southern Africa),      2) the  Channel Islands (Western Europe), 3) Niger (Western Africa),        4)  Qatar (Southwest Asia) and 5) Bhutan (South Asia). Figure \ref{arworld} shows the TFR and LEB functions for these countries. 

TFR in Lesotho has decreased from nearly 6 children in 1960  to 3. The LEB curve reflects a significant problem in Southern Africa: HIV/AIDS. Southern Africa is the worst affected region on the continent, with a very high prevalence (in Lesotho nearly one quarter of the population lives with HIV). We can see how life expectancy was growing from 1960, until the 1990s when the AIDs crisis began. Nowadays LEB in Lesotho is below 50 years. On the other hand, the Channel Islands is representative of countries with low TFR (2.5 in the 1960s and 1.5 now) and high LEB over the years (LEB has risen from 71 to 80 years since the 1960s). On the contrary, Niger is representative of countries with high TFR (around 7) over the years, but low LEB (36 years) in the 1960s, which has increased to nearly 60 years nowadays. TFR in Qatar and Bhutan has decreased spectacularly in this period, from nearly 7 in the 1960s to 2 nowadays. Nevertheless, this decrease has taken place {at} different times. The decline in TFR in Qatar began in the 1970s, whereas in Bhutan it began in 1980s. As regards LEB, in both cases the life expectancy has increased considerably. In the case of Qatar it has increased from 61 years in 1960 to 78 (close to the Channel Islands figure) in 2013. For Bhutan the rise has been more pronounced, since LEB was only 32 years in 1960 (less than that of Niger in 1960), but is now 68 years.

\begin{figure}[H]
\centering
\begin{tabular}{c}
\includegraphics[width=1\textwidth]{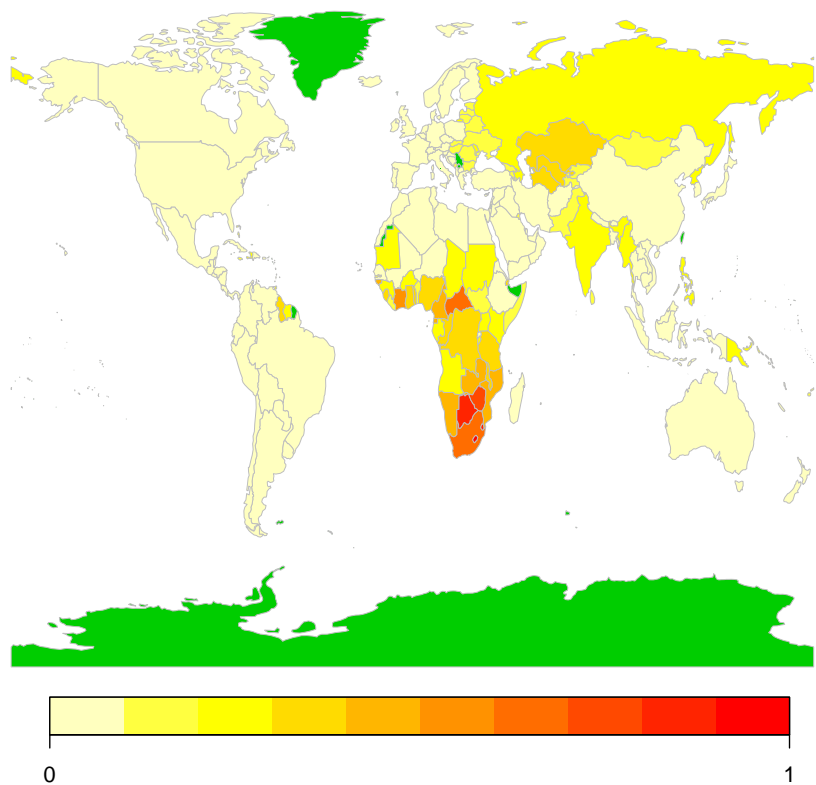}
\end{tabular}
\caption{Visualization of the ${\bf \alpha}$
 coefficients using abundance maps for archetypoid function 1 {(Lesotho, Southern Africa)} of the 5 bivariate (TFR and LEB) archetypoid functions.\label{map51}}
\end{figure}

\begin{figure}[H]
\centering
\begin{tabular}{c}
\includegraphics[width=1\textwidth]{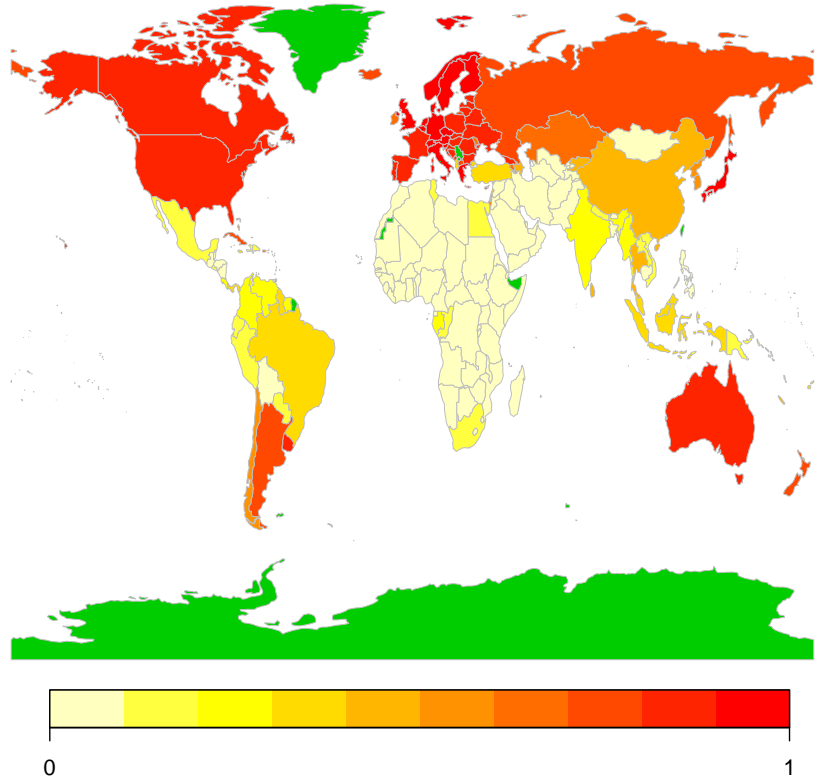}
\end{tabular}
\caption{Visualization of the ${\bf \alpha}$
 coefficients using abundance maps for archetypoid function 2 {(Channel Islands, Western Europe)} of the 5 bivariate (TFR and LEB) archetypoid functions.\label{map52}}
\end{figure}

\begin{figure}[H]
\centering
\begin{tabular}{c}
\includegraphics[width=1\textwidth]{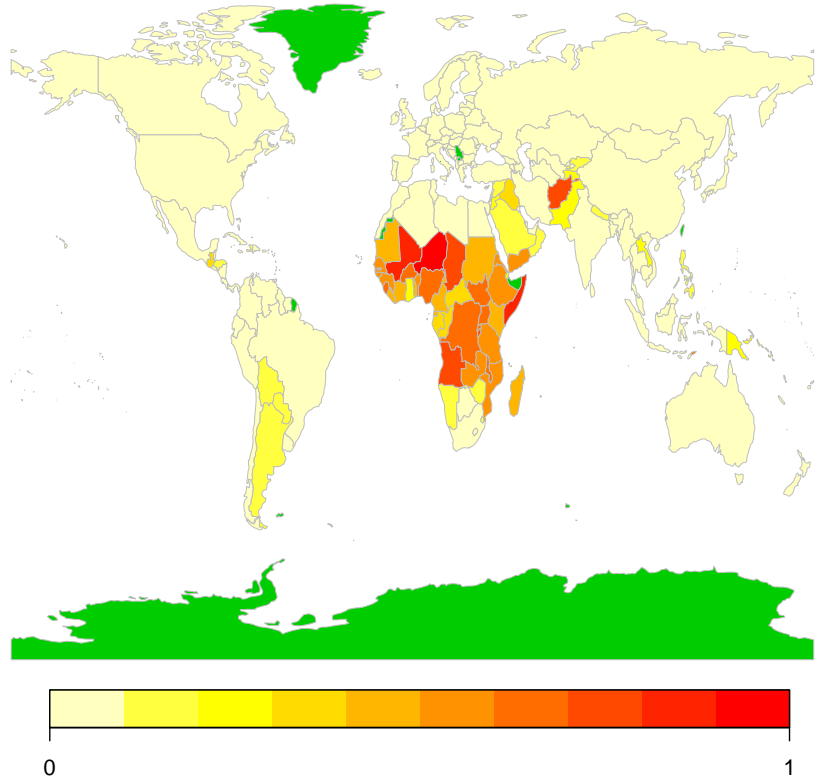}
\end{tabular}
\caption{Visualization of the ${\bf \alpha}$
 coefficients using abundance maps for archetypoid function 3 {(Niger, Western Africa)} of the 5 bivariate (TFR and LEB) archetypoid functions.\label{map53}}
\end{figure}

\begin{figure}[H]
\centering
\begin{tabular}{c}
\includegraphics[width=1\textwidth]{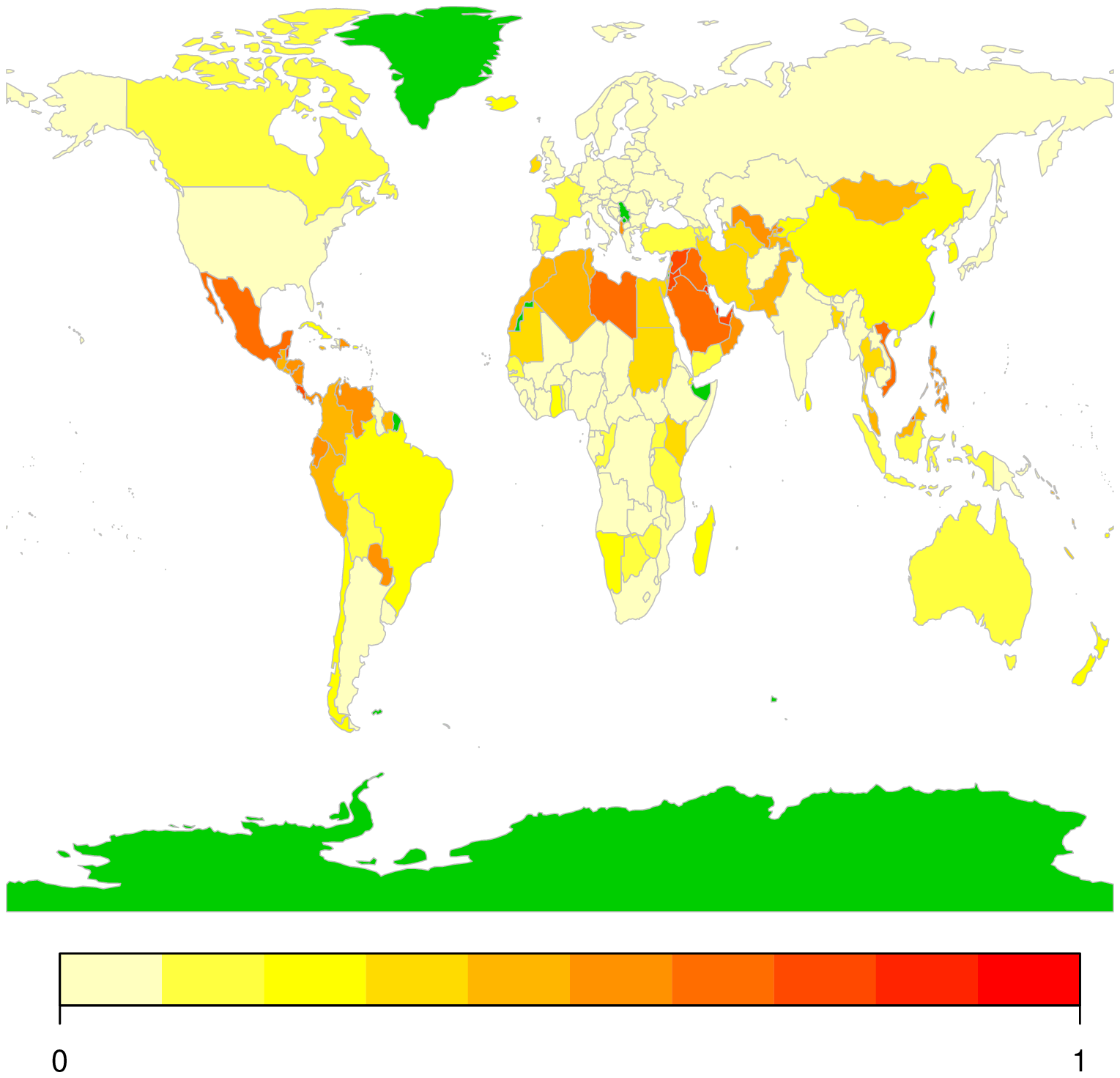}
\end{tabular}
\caption{Visualization of the ${\bf \alpha}$
 coefficients using abundance maps for archetypoid function 4 {(Qatar, Southwest Asia)} of the 5 bivariate (TFR and LEB) archetypoid functions.\label{map54}}
\end{figure}

\begin{figure}[H]
\centering
\begin{tabular}{c}
\includegraphics[width=1\textwidth]{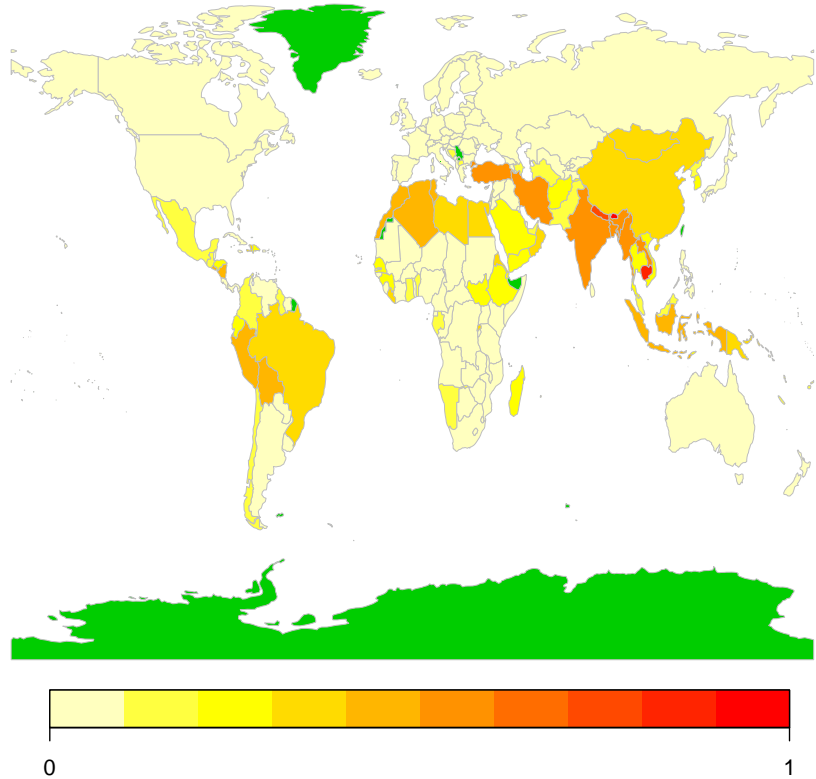}
\end{tabular}
\caption{Visualization of the ${\bf \alpha}$
 coefficients using abundance maps for archetypoid function 5 {(Bhutan, South Asia)} of the 5 bivariate (TFR and LEB) archetypoid functions.\label{map55}}
\end{figure}

Functional archetypoids are representative  of extreme patterns. The interest lies in seeing how the patterns of other countries are expressed in terms of those archetypoids, as a mixture. For that reason, $\bf \alpha$ is computed and represented in the following maps. Each map in Figs. \ref{map51}, \ref{map52}, \ref{map53}, \ref{map54}, \ref{map55} shows the share with the respective functional archetypoid for each country, i.e., it basically shows how well the indicators for each country  are explained by the corresponding functional archetypoid.
 These maps are called abundance maps in the hyperspectral imaging field. The contributions go from 0 to 1. In Figs. \ref{map51}, \ref{map52}, \ref{map53}, \ref{map54}, \ref{map55} these contributions are gradually divided into 10 categories with equally-spaced breaks. The darker the color, the greater the value $\alpha_{ij}$ is, for country $j$ in the functional archetypoid $i$, according to the heat color palette (light yellow indicates low values, whereas red indicates high values). Territories with no information are displayed in green.

According to the maps, we can make the following observations. The countries with indicator curves similar to Lesotho are their neighboring countries, which are the countries most affected by HIV/AIDS. The countries whose indicator functions coincide with those of the Channel Islands are Japan, Australia, North America and most European countries, and to a lesser extent, countries such as Russia and Argentina. With respect to Niger, countries which mainly share  their indicator functions are those in Central Africa and Afghanistan. Countries with a similar behavior to Qatar as regards those indicators are the majority of countries in the Arabian peninsula and neighboring countries, many countries in Central America and several in South America, several in Asia and countries in  North Africa{, a}lthough, those North African countries also share characteristics with Bhutan. {For example, Morocco, Algeria and Tunisia} are a mixture of these two extreme countries with regard to the evolution of those indicators. Other countries are also a mixture of two or three profiles, indicating that they share some of the characteristics of the series. For example, for Turkey the indicators are explained as a mixture between 30\% the Channel Islands, 20\% Qatar and 50\% Bhutan.

%
Although archetypoids are not necessarily nested, in this problem when the number of functional archetypoids {is}  increased, the patterns discovered with smaller $k$ has been kept, and increasing $k$ has led to the discovery of new finer patterns. In other words, although sometimes the archetypoid name is not maintained, and other countries with similar profiles are chosen by the algorithm, the results are nested in this problem.
Table \ref{arkmayor} provides the names of bivariate functional archetypoids for different $k$ values (if different solutions are obtained from the three initial candidates {described in Section \ref{AAmult}}, the one with the lowest RSS is chosen). Countries with similar TFR and LEB series appear in the same row.


\begin{table}[h]
  \centering
  \caption{Functional archetypoids for different $k$ values. $A_j$ denotes the archetypoid number $j$.
 \label{arkmayor}}
 \begin{scriptsize}
  \begin{tabular}{lcccccc}\hline
$k$ = & 5 & 6 & 7 & 8 & 9 & 10\\\hline
$A_1$ & Lesotho & Botswana& Botswana & Botswana & Botswana &Botswana\\
$A_2$& Channel Islands & Channel Islands& Sweden &  Switzerland  &  Switzerland  & Switzerland \\
$A_3$& Niger & Angola & Angola& Niger  & Niger  &Niger \\
$A_4$& Qatar & Jordan & Jordan & Qatar & Qatar & Qatar\\
$A_5$& Bhutan & Bhutan & Bhutan & Maldives & Oman & Oman \\
$A_6$ & & Hong Kong & Hong Kong & Korea, Rep. & Korea, Rep. & Korea, Rep.\\
$A_7$ & & & Russian Federation & Ukraine & Ukraine & Ukraine\\
$A_8$ & & & & Sierra Leone & Sierra Leone &Sierra Leone \\
$A_9$ & & & & &Cambodia &Cambodia \\
$A_{10}$ & & & & & &Rwanda \\
\hline
  \end{tabular}
   \end{scriptsize}
\end{table}

For $k$ = 6, the new pattern incorporated is that corresponding to countries such as Hong Kong SAR, China. {Hong Kong's} TFR has fallen dramatically from 5 in 1960 to even less than 1 or around 1 in recent years (a TFR smaller than that of the Channel Islands). On the other hand, its LEB increased from 67 in 1960 to nearly 84 years in 2013 (a higher LEB  than that of the Channel Islands).
For $k$ = 7, the Russia Federation is included, whose TFR profile  was similar to that of other European countries, although with some key differences. In 1960  TFR in Russia was 2.5 and decreased until 1999, when it was 1.17; it then began to increase until now, and it currently has a TFR of 1.7 (remember that the USSR was dissolved in 1991, with an economic crisis in the 1990s). However, LEB is quite different from most western Europe countries. LEB was 66 years in 1960 and is now 71. Not only  has it not increased very much, but it was below 65 years in 1993 and 1994. With $k$ = 8, 9 and 10, three specific profiles are revealed due to the particularities of these countries, which suffered 
 conflicts that obviously decreased their TFR and LEB very considerably for a time, although at different times: Sierra Leone, Cambodia and Rwanda. In Cambodia the LEB was below 20 years in 1977, and was 26.7 years in Rwanda in 1993.
Solutions are also obtained without considering these three countries {(in case that they are considered as outliers)}, but no very revealing new patterns appear. The same profiles are obtained until $k$ = 7, and for $k$ = 8 the new profile that appears is that corresponding to the Central African Republic. This country had a high $\alpha_{ij}$ corresponding to the Sierra Leone archetypoid, both of which share similar features. {Those three countries did not have influence in the solutions until high $k$, since solutions have not changed after removing them. Note that those countries could exhibit local outlyingness during their war periods, but as functions are examined over an extended period of years and their LEBs were already low, their influence in RSS is not so important for lower $k$. In any data analysis outliers can affect the solutions. This is particularly important in AA, ADA and their functional versions, as we are looking extreme data. Robust AA developed in the classical multivariate case by \cite{Eugster2010} can be extended to the functional case, modifying RSS by using M-estimators instead of least squares estimators.}

\section{Conclusions}
\label{conclusiones}
This paper introduces functional archetype and archetypoid analysis. These techniques can facilitate the understanding of functional data, in the same way that {they do} with {classical} multivariate data. Computational methods are proposed of performing these analyses  based on the coefficients of a basis. Unlike the previous attempt to compute FAA made by \cite{Constatini2012}, which was only valid for an orthogonal basis, the proposed methodology can be used for any basis. It is computationally less demanding than the naive approach of discretizing the functions. Multivariate FAA and FADA are also introduced.

Bivariate FADA has been applied to an interesting problem: the study of human development around the world over the last 50 years. The information contained in two series, TFR and LEB, for each of 190 countries has been shown in maps. These maps  can be easily understood  even by non-experts. The use of FAA and FADA can be an interesting tool for making data easier to interpret, since they are based on the principle of opposites which accommodates human cognition.
With these maps, human development behavior over the last 50 years has been clearly revealed. Figures speak by themselves and tell the story of what has happened in the world in these last 50 years. Firstly, we see the evolution of TFR and LEB. \cite{RoslingTED} explained that ``Swedish top students knew statistically significantly less about the world than chimpanzees''. They thought that nowadays the world was divided into the western world and the third world, i.e.,  small family and long life for the western world and large families with short lives for the third world. However, we have seen that now almost all countries are within the same range (although with a different evolution of curves) with the exception of a few countries. With the maps, the data themselves have also told us about the main historical events, such as a serious health crisis (the HIV pandemic in Sub-Saharan Africa), the breakup of the Soviet Union or 
 wars.


As regards future work, some immediate open problems to address are weighted and robust functional versions, and the definition of AA and ADA for mixed data (functional and vector parts). To work with mixed data, an appropriate interior product could be defined.
Another outstanding issue is to consider FAA and FADA when multivariate arguments are involved. It would be also interesting to explore the sensitivity of the results to the choice of basis. In another vein, instead of AA or ADA, other techniques for non-negative matrix factorization could be extended to the functional case. However, applications will be the main direction of the work. In the same way that AA and ADA have only recently begun  to be used in very diverse fields, FAA and FADA have a very great potential for applications. The most immediate could be along the same lines as the application of this paper, human development. Only TFR and LEB have been considered here, but the number of indicators in the World Bank Open Data is very high (more than 5000), including different and important topics such as Climate Change, Economy, Education, etc., which are of great importance for knowing about development in countries around the globe. As in the case of AA and ADA, FDA applications are growing every day. Both fields are quite new, and therefore there is no doubt plenty of scope for combining them (FAA and FADA).





\section{Acknowledgements}
This work has been partially supported by Grant DPI2013- 47279-C2-1-R. 
 \bibliographystyle{elsarticle-harv}


\end{document}